\documentclass{aa}
\usepackage{txfonts}
\usepackage{graphicx}
\usepackage[latin1]{inputenc}
\def\HII{H\,{\sc{ii}}}

\def\fs{\hbox{$.\!\!^{\rm s}$}}
\def\fdg{\hbox{$.\!\!^\circ$}}
\def\farcm{\hbox{$.\mkern-4mu^\prime$}}
\def\farcs{\hbox{$.\!\!^{\prime\prime}$}}
\def\arcmin{\hbox{$^\prime$}}
\def\arcsec{\hbox{$^{\prime\prime}$}}

\def\sun{\hbox{$\odot$}}
\def\degr{\hbox{$^\circ$}}
\def\h{\hbox{$^{\reset@font\r@mn{h}}$}}
\def\m{\hbox{$^{\reset@font\r@mn{m}}$}}
\def\s{\hbox{$^{\reset@font\r@mn{s}}$}}
\def\msol{\hbox{\kern 0.20em $M_\odot$}}

\def\kms{\hbox{\kern 0.20em km\kern 0.20em s$^{-1}$}}
\def\cmmt{\hbox{\kern 0.20em cm$^{-3}$}}
\def\cmmd{\hbox{\kern 0.20em cm$^{-2}$}}
\def\pc{\hbox{\kern 0.20em pc$^{2}$}}

\def\h13cop{\hbox{H$^{13}$CO$^{+}$}}

\begin{document}
\title{Triggered massive-star formation \\
on the borders of Galactic \HII\ regions
\thanks{Based on observations obtained at the European Southern
Observatory using the ESO New Technology Telescope (NTT) (program
70.C-0296) and the ESO Swedish Submillimetre Telescope (program
71.A-0566), on La Silla, Chile}}
 \subtitle{II. Evidence for the
collect and collapse process around RCW~79}

\author{A.~Zavagno\inst{1}
\and L.~Deharveng\inst{1}
\and
      F. Comer{\'o}n\inst{2}
 \and J.~Brand\inst{3}
         \and
       F. Massi\inst{4}
       \and
         J.~Caplan\inst{1}
         \and
         D.~Russeil\inst{1}
 }
  \authorrunning{A. Zavagno et al.}
  \offprints{A. Zavagno}

\institute{
      Laboratoire d'Astrophysique de Marseille, 2 place Le Verrier, 13248 Marseille Cedex 4, France
             \and
        European Southern Observatory, Karl-Schwarzschild-Strasse 2, D-85748 Garching,
        Germany \and
      INAF-Istituto di Radioastronomia, Via Gobetti 101, 40129 Bologna, Italy
        \and
        INAF-Osservatorio Astrofisico di Arcetri, Largo E. Fermi, 5, 50125 Firenze, Italy
  }
\date{Received 2005-07-30; accepted 2005-09-07 }

\abstract{We present SEST-SIMBA 1.2-mm continuum maps and ESO-NTT
SOFI $JHK_{S}$ images of the Galactic \HII\ region RCW~79. The
millimetre continuum data reveal the presence of massive fragments
located in a dust emission ring surrounding the ionized gas. The two
most massive fragments are diametrically opposite each other in the
ring. The near-IR data, centred on the compact \HII\ region located
at the south-eastern border of RCW~79, show the presence of an
IR-bright cluster containing massive stars along with young stellar
objects with near-IR excesses. A bright near- and mid-IR source is
detected towards maser emissions, 1.2~pc north-east of the compact
\HII\ region centre. Additional information, extracted from the
Spitzer GLIMPSE survey, are used to discuss the nature of the bright
IR sources observed towards RCW~79. Twelve luminous Class~I sources
are identified towards the most massive millimetre fragments. All
these facts strongly indicate that the massive-star formation
observed at the border of the \HII\ region RCW~79 has been triggered
by its expansion, most probably by the collect and collapse process.
   \keywords{Stars: formation -- Stars: early-type -- ISM: \HII\ regions --
   ISM: individual: RCW~79}
}
 \titlerunning{Triggered massive-star formation}
 \authorrunning{A. Zavagno et al.}

\maketitle
\noindent
%_____________________
\section{Introduction \label{intro}}

Several physical processes linked to the expansion of \HII\ regions
may trigger star formation. A review of the different processes is
given by Elmegreen (\cite{elm98}). Among these processes, we are
interested in the collect and collapse mechanism  because it leads
to the formation of massive objects (stars or clusters, Deharveng et
al. \cite{deh05}, hereafter paper I). This process, first proposed
by Elmegreen \& Lada (\cite{elm77}), has been treated analytically
by Whitworth et al. (\cite{whi94}). Because of the supersonic
expansion of an \HII\ region into the surrounding medium, a
compressed layer of gas and dust accumulates between the ionization
and the shock fronts. With time this layer grows in mass and
possibly becomes gravitationally unstable, fragments, and forms
massive cores. Those cores represent potential sites of
second-generation massive-star formation.

To better understand this process of triggered massive-star
formation, we have begun a multi-wavelength study of the borders of
Galactic \HII\ regions. We selected \HII\ regions with a simple
morphology (circular ionized regions surrounded by dust emission
rings in the mid-IR) and hosting signposts of massive-star formation
at their peripheries (luminous IR sources, ultracompact radio
sources; see paper I for details about the selection criteria and
the sample). Near-IR imaging gives information about
 the stellar nature of the luminous IR point
sources observed at the borders of the ionized regions. It allows us
to identify massive stars there and hence to confirm that massive
objects (stars or clusters) are indeed formed via this process.
Millimetre data (molecular emission lines or/and dust emission
continuum) are used to search for massive fragments along an annular
structure surrounding the ionized gas. The IR sources should be
observed inside or close to these molecular cores. Sh~104, a
Galactic \HII\ region, is the prototype of \HII\ regions
experiencing the collect and collapse process to form massive stars
(Deharveng et al. \cite{deh03}). Other candidates proposed in
paper~I are under analysis; RCW~79 is one of these. It has been
studied in detail by Cohen et al.~(\cite{coh02}). The presence of a
compact \HII\ region located at its south-east border, just behind
its ionization front, indicates that massive-star formation has
taken place there. Cohen et al. proposed that RCW~79 encountered a
massive molecular clump during its expansion, triggering star
formation there (see the end of their sect.~5). We present in this
paper new observational facts, based on near-IR, mid-IR and
millimetre continuum data, that support the hypothesis of the
collect and collapse mechanism being at work there. Sect.~2
introduces the RCW~79 region. Observations and data reduction are
presented in Sect.~3. The results are presented in Sect.~4 and are
discussed in Sect.~5. Our conclusions are given in Sect.~6.
%_______________________________________________________________________
\section{Presentation of RCW~79}
%-----------------------------------------------------------------------
RCW~79 (Rodgers, Campbell and Whiteoak~\cite{rod60}, $l$=308\fdg6,
$b$=0\fdg6) is a bright optical \HII\ region of diameter
$\sim$12\arcmin, located at a distance of 4.3~kpc (Russeil~\cite{rus03}). This
region and its surroundings have been studied by Cohen et
al.~(\cite{coh02}), hereafter CGPMC. CGPMC present a 843~MHz
continuum emission map showing a shell nebula, corresponding to the
optical \HII\ region, and a compact \HII\ region, without any
optical counterpart, at the south-east border of the nebula's shell.
The velocity of the ionized gas is in the $-40$ to $-51$~km~s$^{-1}$
range (see Sect.~\ref{vha}).

\begin{figure}[tb]
  \includegraphics[angle=0,width=85mm ]{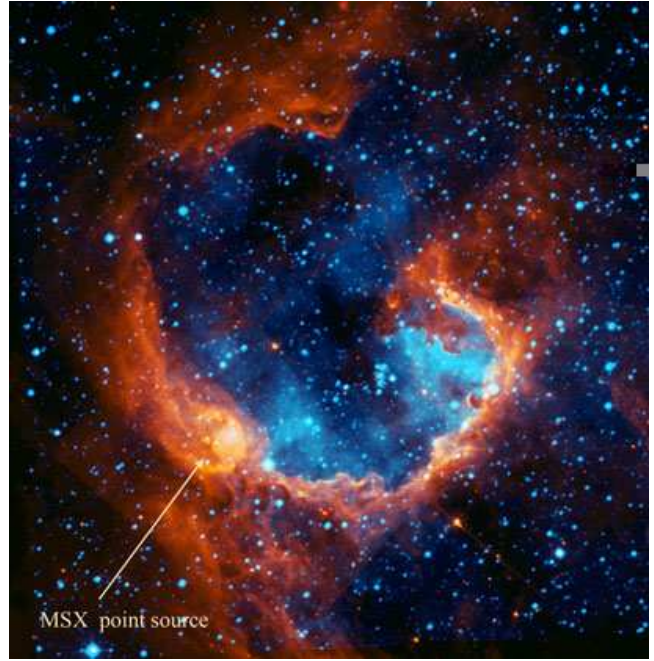}
  \caption{Spitzer-IRAC 8\,$\mu$m image from the GLIMPSE survey (orange) superimposed on a SuperCOSMOS H$\alpha$ image (turquoise)
  of RCW~79. The location of the MSX point source is indicated. The field size is 15\arcmin$\times$15\arcmin  }
  \label{haspitzer}
\end{figure}

RCW~79 is nearly completely surrounded by a dust ring (see
Fig.~\ref{haspitzer}), revealed by its
mid-IR emission in the MSX Band A (centred at 8.3\,$\mu$m, Price et
al. \cite{pri01}) and in the Spitzer IRAC (Fazio et
al.~\cite{faz04}) channel 4 (centered at 8\,$\mu$m, GLIMPSE survey,
Benjamin et al.~\cite{ben03}). In the hot photodissociation regions
(PDRs) surrounding \HII\ regions, the emission in the MSX Band A and
in IRAC (channel 4 is dominated by emission bands centred at 7.7 and
8.6\,$\mu$m and commonly attributed to polycyclic aromatic
hydrocarbon-like molecules, PAHs; L\'eger \& Puget \cite{leg84}).
Fig.~\ref{haspitzer} presents a colour composite image of this
region. The H$\alpha$ emission of the ionized gas (from the
SuperCOSMOS H-alpha survey, Parker \& Phillips~\cite{par98}) appears
in turquoise. The dust emission (image from the Spitzer GLIMPSE
survey) in the band centered at 8\,$\mu$m appears in orange. These
two emissions are clearly anti-correlated. PAHs are destroyed in the
ionized region, but are present in the photo-dissociated region,
where they are excited by the photons leaking from the \HII\ region.

%-------------------------
% Table~1
\begin{table*}
\caption{Coordinates of objects discussed in the text}
\begin{tabular}{lll}
 \hline\hline
 Object &  \multicolumn{1}{c}{$\alpha_{2000}$} & \multicolumn{1}{c}{$\delta_{2000}$}  \\
% Object    & RA (2000) & &     & Dec (2000) & &      \\
  \hline
 Centre of RCW~79  &  13$^{\rm h}$ 40$^{\rm m}$ 17$\fs$0 &
$-$61$\degr$  44$\arcmin$ 00$\arcsec$ \\
IRAS~13374$-$6130 & 13$^{\rm h}$ 40$^{\rm m}$  53$\fs$2 & $-$61$\degr$  45$\arcmin$  47$\arcsec$ \\
 MSX point source  &  13$^{\rm h}$ 40$^{\rm m}$  53$\fs$1 & $-$61$\degr$  45$\arcmin$  51$\arcsec$ \\
 Centre of the compact \HII\ region$^{\footnotesize 1}$  & 13$^{\rm h}$ 40$^{\rm m}$  52$\fs$5 & $-$61$\degr$  45$\arcmin$  54$\arcsec$\\
 Maser sources      &  13$^{\rm h}$ 40$^{\rm m}$  57$\fs$6 & $-$61$\degr$  45$\arcmin$ 43$\arcsec$ \\
 \hline
\end{tabular} \\
\\
$^{\footnotesize 1}$ Not given in CGPMC. Measured from their figs.~3
and 5 (about 5\arcsec\, accuracy)\\

\end{table*}

Table~1 gives the position of sources discussed in the text. The IR
source G308.7452+00.5482 in the MSX Point Source Catalog (Egan et
al. \cite{ega99}) lies in the direction of the dust ring, towards the compact \HII\ region
at the south-eastern border of RCW~79 (see Fig.~\ref{haspitzer}). This
point source also corresponds to IRAS~13374$-$6130. It has the
luminosity (55000 $L_{\sun}$) and colours of a UC \HII\ region
(CGPMC, paper I). The high angular resolution of the GLIMPSE survey
allows us to resolve the MSX point source. Fig.~\ref{haspitzer}
shows that it is composed of several compact components surrounded
by a small dust emission ring of diameter $1\farcm 7$. The compact
\HII\ region is not seen on the H$\alpha$ image, indicating a high
visual extinction in this direction. The near-IR data presented in
this paper focus on this compact \HII\ region where they unveil the
presence of a small cluster.

RCW~79 belongs to a molecular complex, in Centaurus, studied by
Saito et al.~(\cite{sai01}) using the NANTEN telescope (see their
figs.~1a and 4). The $^{12}$CO (1--0) emission, integrated over the
velocity range $-64$ to $-36$~km~s$^{-1}$, shows three condensations
at the borders of RCW~79. The  C$^{18}$O (1--0) emission integrated
over the $-$50 to $-$44 km s$^{-1}$ range shows a massive
condensation ($M\simeq 6000$--$12000 M_{\odot}$) in the direction of
the compact \HII\ region. The mean H$_2$ density in this
condensation is 1400~cm$^{-3}$.

Maser emissions (OH, methanol and water) have been detected
in RCW~79 towards the mid-IR MSX point source (Caswell \cite{cas04},
van der Walt et al. \cite{van95}; see also Table~1), indicating that
massive-star formation is taking place. The emission peak of the
methanol and water masers are observed at $-51$ and $-$49 km
s$^{-1}$, respectively, indicating that those emissions are indeed
associated with RCW~79.

Fig.~\ref{haspitzer} shows a hole in the ring of dust emission
surrounding RCW~79, to the north-west. H$\alpha$ emission is
observed in this direction, suggesting that RCW~79 may be
experiencing a champagne phase, the ionized gas flowing away from
the \HII\ region through this hole. We discuss this point in
Sect.~\ref{vha}.
\subsection{Radio continuum emission \label{rad}}
The MGPS2 (Green \cite{gre99}) 843~MHz radio continuum map presented
by CGPMC (their fig.~5), obtained with a resolution of $43\arcsec
\times 49\arcsec$, shows that the radio source corresponding to
RCW~79 consists of a large shell nebula, of diameter $\sim
12\arcmin$ (similar to the optical nebula), and a compact region. No
flux density has been published for this source. From the map we
estimate that its flux density is about 1/16 that of the total
source. Assuming that we are dealing with thermal emission, and
adopting the total flux density of the source, 17~Jy at 5~GHz,
measured by Caswell \& Haynes (\cite{cas87}), we obtain fluxes of
1~Jy and 16~Jy respectively for the compact and extended radio
sources.

These radio flux densities allow us to estimate the ionizing-photon
fluxes, and hence the spectral types of the main exciting stars
(assuming that a single main-sequence star dominates the ionization
in each region). Using eq.~1 of Simpson \& Rubin (\cite{sim90}), and
a distance of 4.3~kpc, we derive ionizing fluxes $N_{\mathrm{Lyc}}$
of $2.9\times10^{49}$ and $1.9\times10^{48}$ photons~s$^{-1}$ for
the extended and the compact \HII\ regions, respectively. There are
large uncertainties in the effective temperatures and the ionizing
fluxes of massive stars, for a given spectral type. According to
Vacca et al. (\cite{vac96}) and to Schaerer \& de Koter
(\cite{sch97}), these ionizing fluxes correspond to O5V and O9.5V
stars, respectively. According to Smith et al. (\cite{smi02}), these
fluxes correspond to O3V--O4V and O8.5V stars, respectively.
According to Martins et al. (\cite{mar05}) they indicate O4V and O8V
stars. Thus the large RCW~79 \HII\ region is a high-excitation
region possibly ionized by a star in the O3V--O5V range, while the
compact \HII\ region is of lower excitation, possibly ionized by a
star of spectral type O8V--O9.5V. On the other hand, the IR
luminosity of the IRAS point source observed in the direction of the
compact \HII\ region (55000 $L_{\odot}$) gives a limit of O9V for
the most luminous/massive star of the exciting cluster. New radio
continuum maps at higher resolution are needed to resolve the
compact \HII\ region and measure its size and radio flux.

\section{Observations}
\subsection{SEST-SIMBA 1.2-millimetre continuum imaging \label{mmobs}}
Continuum maps at 1.2-mm (250~GHz) of a 20\arcmin $\times$
20\arcmin\, field centred on RCW~79 have been obtained using the 37
channel SIMBA bolometer array (SEST Imaging Bolometer Array) mounted
at SEST on May 7-8 2003. The beam size is 24\arcsec. Nine
individuals maps covering the whole region were obtained with the
fast scanning speed (80\arcsec per second). The total integration
time was 10.5 hours. The final map was constructed by combining the
individual maps. Skydips were performed after each integration to
determine the atmospheric opacity. Maps of Uranus and Neptune were
obtained for the calibration. The individual maps were reduced and
analyzed using MOPSIC, a software package developed by Robert Zylka
(Grenoble Observatory ; see also
http://www.astro.ruhr-uni-bochum.de/nielbock/simba/mopsic.pdf). The
common procedures are described in Chapter 4 of the SIMBA Observers
Handbook. A detailed description of the data reduction can be found
in Chini et al. (\cite{chi03}).

The data reduction of the nine individual maps is done in two steps.
The first step includes a global baseline fit, despiking,
deconvolution of the instrumental bandpass, gain-elevation, opacity
corrections and skynoise reduction. A map is then created by
combination of individual maps. This map is used to define a polygon
that includes the emission zone. Then the same procedure is applied
to the original FITS files but using the zone outside the polygon
for the baseline fitting. The final map is obtained at the end of
this second iteration. The calibration was obtained using planet
maps. We derived a conversion factor of 55~mJy/count for the first
day and 51~mJy/count for the second day, applied respectively to the
individual maps before the final combining. The residual noise in
the final map is about 20~mJy/beam (1$\sigma$). We then used the
emission above 5$\sigma$ (100 mJy/beam level) to define the 1.2-mm
condensations.

\subsection{ESO-NTT near-IR imaging \label{nirobs}}
Images in the $J$ (1.25~$\mu$m), $H$ (1.65~$\mu$m), and $K_S$
(2.2~$\mu$m) near-infrared bands  were obtained on the nights of 14
and 15 February 2003 with the SofI camera at the ESO New Technology
Telescope (NTT). They cover a $2\farcm66\times3\farcm85$ area
centered on the bright MSX point source. The corresponding images
obtained on the first night combined 40 ($J$), 19 ($H$), and 10
($K_S$) separate frames obtained with small offsets in between,
using stellar images to determine the telescope offsets before
combination. Each frame consisted in turn of 20 individual exposures
co-added on the detector, with individual integration times of 2.4,
2.4, and 1.2~sec in the $J$, $H$ and $K_{S}$ filters. The rather
short detector integration times were chosen because of the presence
of bright saturating stars in the field of view. The observation
sequence was repeated on the following night, this time obtaining
respectively 26, 24, and 8 frames with 15 individual exposures of 4,
2.5, and 2~sec. The individual images were dark-subtracted, divided
by a master flat field, and sky-subtracted before combination. Both
the amplitude of the telescope offset pattern and the number of
exposures in each filter were large enough to allow us to produce an
acceptable sky frame by median averaging the dark-subtracted,
flat-fielded frames uncorrected for the telescope motion, clipping
off in the median averaging the upper half of the pixel values at
each detector position.

Sources were detected using DAOFIND (Stetson~\cite{ste87}). A master
list of sources was produced by running DAOPHOT on a single image
built by adding the $J$, $H$, and $K_S$ images, so as to ensure that
all sources observed in at least one filter entered this master
list. Unsaturated and relatively isolated bright stars in the
co-added $JHK_S$ image were used to determine an approximate PSF,
needed for automated point source identification. Photometry was
then performed on the images obtained through each filter by
defining an undersized aperture at the position of each star in the
master list, measuring the flux inside it, and adding the rest of
the flux in the PSF as given by the fit of a circularly symmetric
radial profile to each stellar image. This procedure allowed both to
reduce the contamination to the aperture photometry by other stars
located on the wings of the PSF, and to adjust to the mildly
variable image quality across the field of view. Finally, zero-point
calibration was carried out by observing six infrared standard stars
from Persson et al. (\cite{per98}) at different air-masses during
both nights.

\subsection{H$\alpha$ observations}
H$\alpha$ Fabry-Perot observations of RCW~79 were obtained with the
CIGALE instrument on a 36-cm telescope (La Silla, ESO) in May 1990.
A description of the instrument, including data acquisition and
reduction techniques, can be found in le Coarer et
al.~(\cite{lec92}). The field of view is 39$\arcmin$ with a pixel
size of 9$\arcsec$. The Fabry-Perot interferometer has an
interference order of 2604 at H$\alpha$, providing a free spectral
range equivalent to 115~km~s$^{-1}$, and a spectral sampling of 5~
km~s$^{-1}$. The velocity accuracy is $\sim$1~km~s$^{-1}$.

The H$\alpha$ profiles are decomposed into several components:
H$\alpha$ geocoronal and OH night-sky lines, and nebular lines. The
channel maps show that the nebular velocity ranges from
$-$72~km~s$^{-1}$ to $-$18~km~s$^{-1}$. The H$\alpha$ emission
associated with RCW~79 is in the range $-$51 to $-$40~km~s$^{-1}$.
Two other H$\alpha$ emissions are superimposed along the line of
sight: the local arm emission at V$_{LSR} \sim 0$~km~s$^{-1}$, and
that of the Sagittarius arm at V$_{LSR} \sim -25$~km~s$^{-1}$. To increase
the S/N ratio, profiles were binned over 5~pixel$\times$5~pixel
areas.

\section{Results}
%________________
\subsection{Continuum imaging at 1.2-mm \label{mmres}}
%____________________________________________________

\begin{figure}[tb]
  \includegraphics[angle=270,width=85mm ]{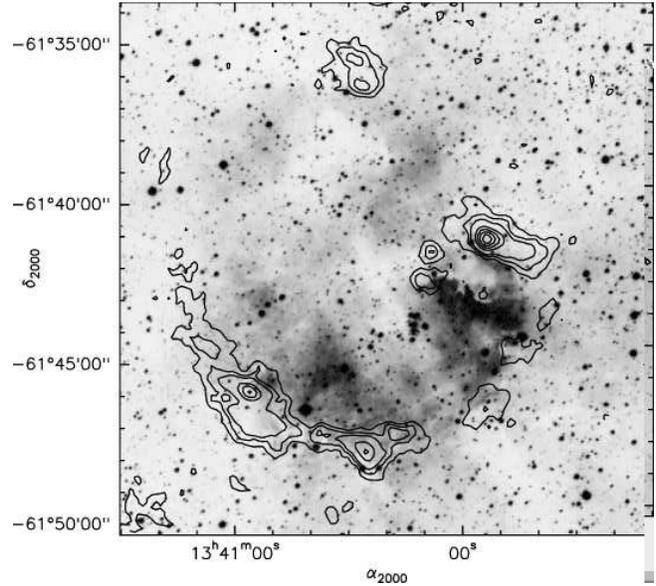}
  \caption{Millimetre continuum emission (contours) superimposed on a SuperCOSMOS H$\alpha$ image of the region.
  The first levels are 50, 100 and 150 and then increase in steps of 100 mJy/beam, from 300 to 700~mJy/beam}
  \label{hamm}
\end{figure}

Figure~\ref{hamm} shows the 1.2-mm continuum emission superimposed
on a SuperCOSMOS H$\alpha$ image of the region.
\begin{figure}[tb]
  \includegraphics[angle=0,width=85mm ]{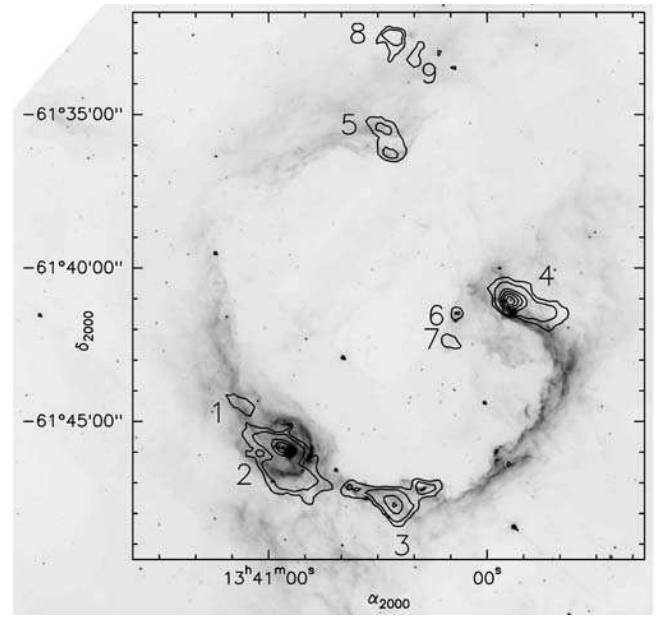}
  \caption{Millimetre continuum emission (contours) superimposed on the GLIMPSE 8\,$\mu$m image of the region.
  The levels are 100, 150~mJy/beam and then increase from 300 to 700~mJy/beam in steps of 100~mJy/beam. The condensations discussed in the text
  are identified by their numbers (from 1 to 9). Their limits, used for the mass derivations, are defined by the 100~mJy/beam contour (5$\sigma$ level, see text)}
\label{spitzermm}
\end{figure}
The fact that this 1.2-mm emission surrounds the ionized region, is
located just beyond the ionization front (Fig.~\ref{hamm}) and shows
the same annular structure as shown by the mid-IR dust emission
(Figs.~\ref{haspitzer} and \ref{spitzermm}) demonstrates that this
emission is associated with RCW~79. Fig.~\ref{spitzermm} presents
the 1.2-mm continuum emission (as contours) superimposed on the
8\,$\mu$m GLIMPSE image of RCW~79. The lowest contour shown in
Fig.~\ref{spitzermm} delineates the condensations' surface as
defined, for the mass estimates, at the 5$\sigma$ level
(100~mJy/beam, Sect.~\ref{mmobs}).

The millimetre continuum map reveals the presence of nine fragments.
Five (nos 1 to 5) of the nine form an emission ring surrounding the
main ionized region RCW~79. Two (nos 6 and 7) are situated in front
of the ionized gas, as shown by the absorption features observed
in the visible (Fig.~\ref{hamm}).
Fragments 8 and 9, in the North, are most probably related to the
region, judging from the faint GLIMPSE mid-IR emission
observed there.

\subsubsection{Mass estimates \label{massest}}
The millimetre continuum emission from condensations identified in
Fig.~\ref{spitzermm} is mainly due to optically thin thermal dust
emission. Following Hildebrand (\cite{hil83}) and assuming standard
dust properties and gas-to-dust ratio, the integrated millimetre
flux is related to the total (gas+dust) mass of the condensations.
For the mass estimates, we used the 1.2-mm integrated fluxes,
$F^{\mathrm{int}}_{\mathrm{1.2mm}}$, given in Table~\ref{massemm}.
For condensation 2, we did not correct this flux for the free-free
emission of the compact \HII\ region. Indeed, this emission does not
coincide with the millimetre emission peak, and higher resolution
radio data are needed to accurately estimate its contribution. Note that the dust
mass estimate for this condensation is, therefore, an upper limit.

According to Hildebrand (\cite{hil83}) the ``millimetre'' (gas+dust) mass is
related to the flux by
\begin{eqnarray*}
M_{\mathrm{(gas+dust)}}\, (M_{\odot}) = g\,\,\frac{F^{\mathrm{int}}_{\mathrm{1.2mm}}
(\mathrm{Jy}) \,\, (D\, (\mathrm{kpc}))^2}{\kappa_{\mathrm{1.2mm}}\,\,
B_{\mathrm{1.2mm}}(T_{\mathrm{dust}})},
\end{eqnarray*}
where $\kappa_{\mathrm{1.2mm}}$ is the dust opacity per unit mass at
1.2-mm and $B_{\mathrm{1.2mm}}(T_{\mathrm{dust}})$ the Planck
function for a temperature $T_{\mathrm{dust}}$. The values of
$T_{\mathrm{dust}}$ and $\kappa_{\mathrm{1.2mm}}$ are unknown and
have to be assumed to derive the masses. The dust temperature should
be in the 20--30~K range, as usually assumed for protostellar
condensations (cf.\ Motte et al. \cite{mot03}). Ossenkopf \& Henning
(\cite{oss94}) recommended using a value of
$\kappa_{\mathrm{1.2mm}}$~=~1 cm$^{2}$~g$^{-1}$ for a protostellar
cores dust mass and $g$ is the gas-to-dust ratio that we assume to be
100. Table~\ref{massemm}
lists the measured and derived properties obtained for the
millimetre fragments identified in Fig.~\ref{spitzermm}. Column 1
gives the fragment numbers, columns 2 and 3 give the emission peak
coordinates, column 4 gives the 1.2-mm integrated flux, and column~5
the range of derived masses for the corresponding fragment,
depending on the adopted temperature (20 or 30~K). Note that the
lower mass values correspond to the higher dust temperature.
\begin{table*}
\caption{Mass estimates for the millimetre fragments}
\begin{tabular}{c c c c c}

\hline\hline
 Number & \multicolumn{2}{c}{Peak position} & $F^{\mathrm{int}}_{\mathrm{1.2mm}}$$^{\footnotesize 1}$ & Mass range$^{\footnotesize 2}$ \\
        &            $\alpha_{2000}$ & $\delta_{2000}$   & (mJy)              & ($M_{\odot}$) \\
  \hline
        &            &     &                  &                   \\
  1 & 13$^{\rm h}$ 40$^{\rm m}$ 56$\fs$00 &  $-$61$\degr$ 45$\arcmin$ 51$\farcs$75  & 224 & 42 -- 70    \\
  2 & 13$^{\rm h}$ 41$^{\rm m}$ 09$\fs$30 &  $-$61$\degr$ 44$\arcmin$ 22$\farcs$76 & 4200 & 796 -- 1326$^{\footnotesize 3}$    \\
  3 & 13$^{\rm h}$ 40$^{\rm m}$ 25$\fs$45 &  $-$61$\degr$ 47$\arcmin$ 43$\farcs$68 & 2350 & 444 -- 742 \\
  4 & 13$^{\rm h}$ 39$^{\rm m}$ 54$\fs$00 &  $-$61$\degr$ 41$\arcmin$ 04$\farcs$22 & 3000   & 567 -- 947 \\
  5 & 13$^{\rm h}$ 40$^{\rm m}$ 26$\fs$75 &  $-$61$\degr$ 36$\arcmin$ 16$\farcs$10  & 795 & 150 -- 251\\
  6 & 13$^{\rm h}$ 40$^{\rm m}$ 08$\fs$11 &  $-$61$\degr$ 41$\arcmin$ 27$\farcs$87  & 81 & 15 -- 25 \\
  7 & 13$^{\rm h}$ 40$^{\rm m}$ 10$\fs$58 &  $-$61$\degr$ 42$\arcmin$ 23$\farcs$85   & 144 & 27 -- 45  \\
  8 & 13$^{\rm h}$ 40$^{\rm m}$ 25$\fs$10 &  $-$61$\degr$ 32$\arcmin$ 27$\farcs$05   & 483 & 91 -- 153 \\
  9 & 13$^{\rm h}$ 40$^{\rm m}$ 19$\fs$90 &  $-$61$\degr$ 33$\arcmin$ 13$\farcs$40   & 173 & 33 -- 55 \\
  \hline
  \label{massemm}
\end{tabular}\\
\\
{$^{\footnotesize 1}$ 1.2-mm flux integrated above the 5$\sigma$ level} \\
{$^{\footnotesize 2}$ Mass calculations performed with $T_{\rm{dust}}$=30 and 20~K for the lower and higher values, respectively}   \\
{$^{\footnotesize 3}$ Upper limit (see text)}
\\
\\
%{$^{\footnotesize 1}$ Corrected from the free-free contribution (see text)}   \\
\end{table*}

\subsection{Stellar content of the millimetre condensations \label{spitzer}}
%---------------------------------------------------------------------------
We have investigated the  stellar content of the three most massive
millimetre condensations (nos 2, 3 and 4, see Fig.~\ref{spitzermm}).
We are particularly interested in identifying red and luminous
objects observed in the direction of these condensations. Such
sources may represent embedded massive young stellar objects, whose formation
has been triggered by the expansion of the RCW~79 \HII\ region. We
look in more detail at condensation 2 where the stellar cluster
ionizing the compact \HII\ region is observed. The near-IR ESO-NTT
observations cover this region.

\begin{figure}[tb]
  \includegraphics[angle=0,width=85mm ]{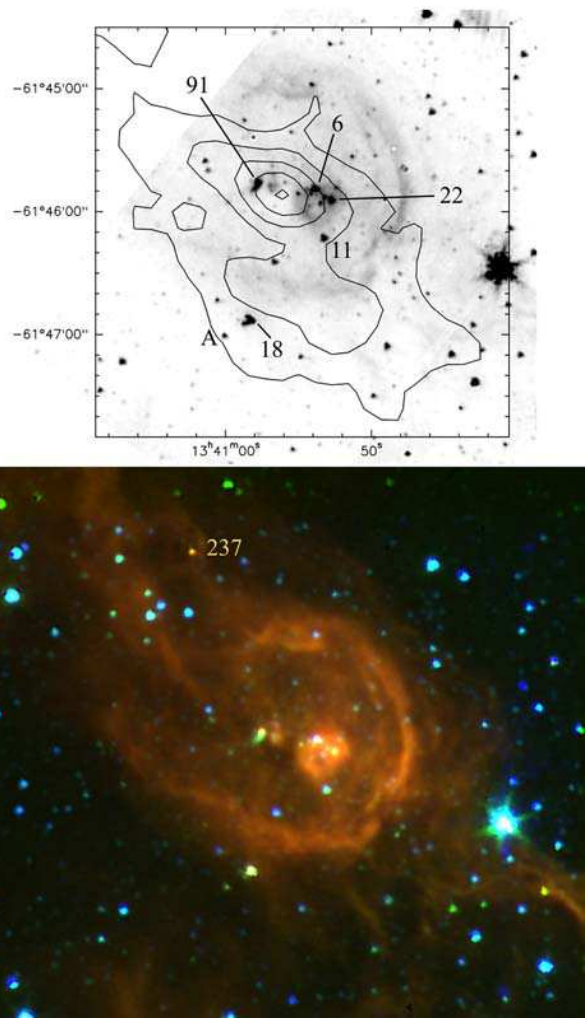}
  \caption{Condensation~2 - Top) The 1.2-mm continuum emission is superimposed, in
  contours, on the 3.6\,$\mu$m image taken from the Spitzer
  GLIMPSE survey. Bottom) Color composite image combining the $K_{S}$ image from the 2MASS survey
  (in blue), and the 3.6\,$\mu$m (in green) and the 8\,$\mu$m (in red), both from the GLIMPSE survey.
  The stars discussed in the text are identified by their numbers in Table~3}
  \label{cond2smmf}
\end{figure}

\begin{figure}[tb]
  \includegraphics[angle=0,width=85mm ]{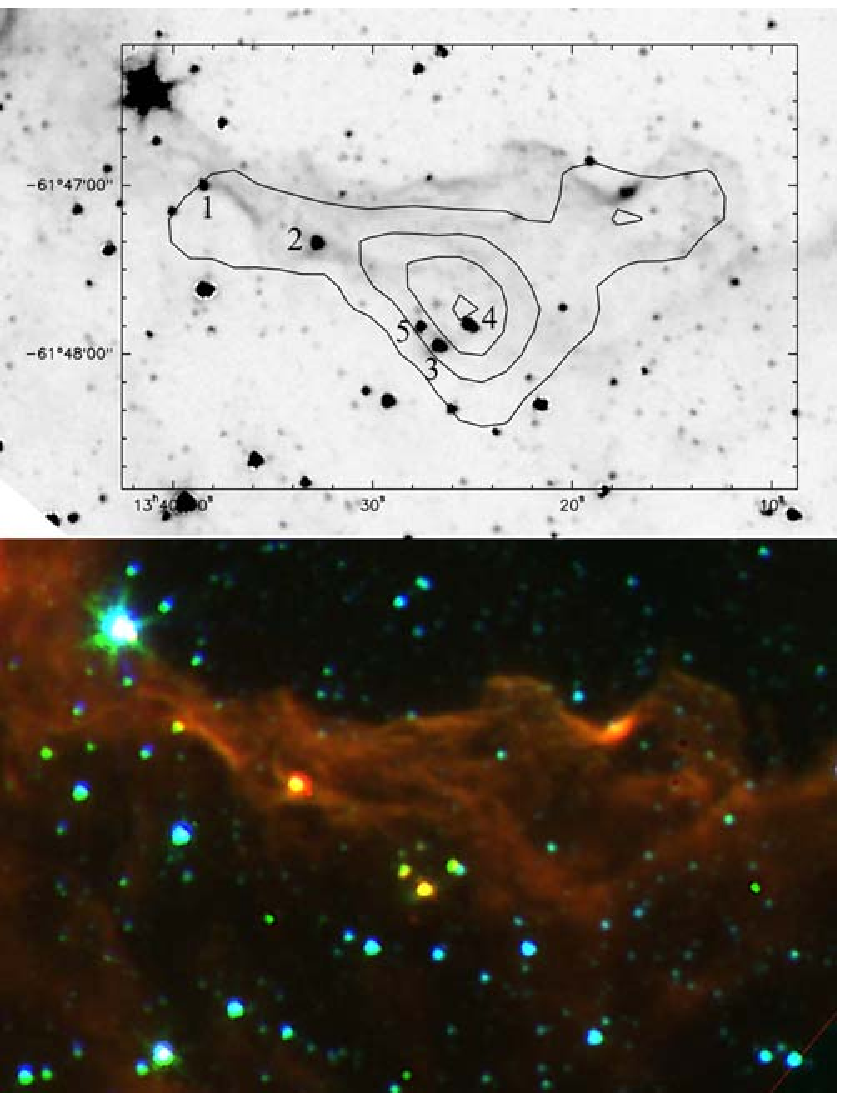}
  \caption{As Fig.~\ref{cond2smmf} but for condensation 3}
  \label{cond3smmf}
\end{figure}

\begin{figure}[tb]
  \includegraphics[angle=0,width=85mm ]{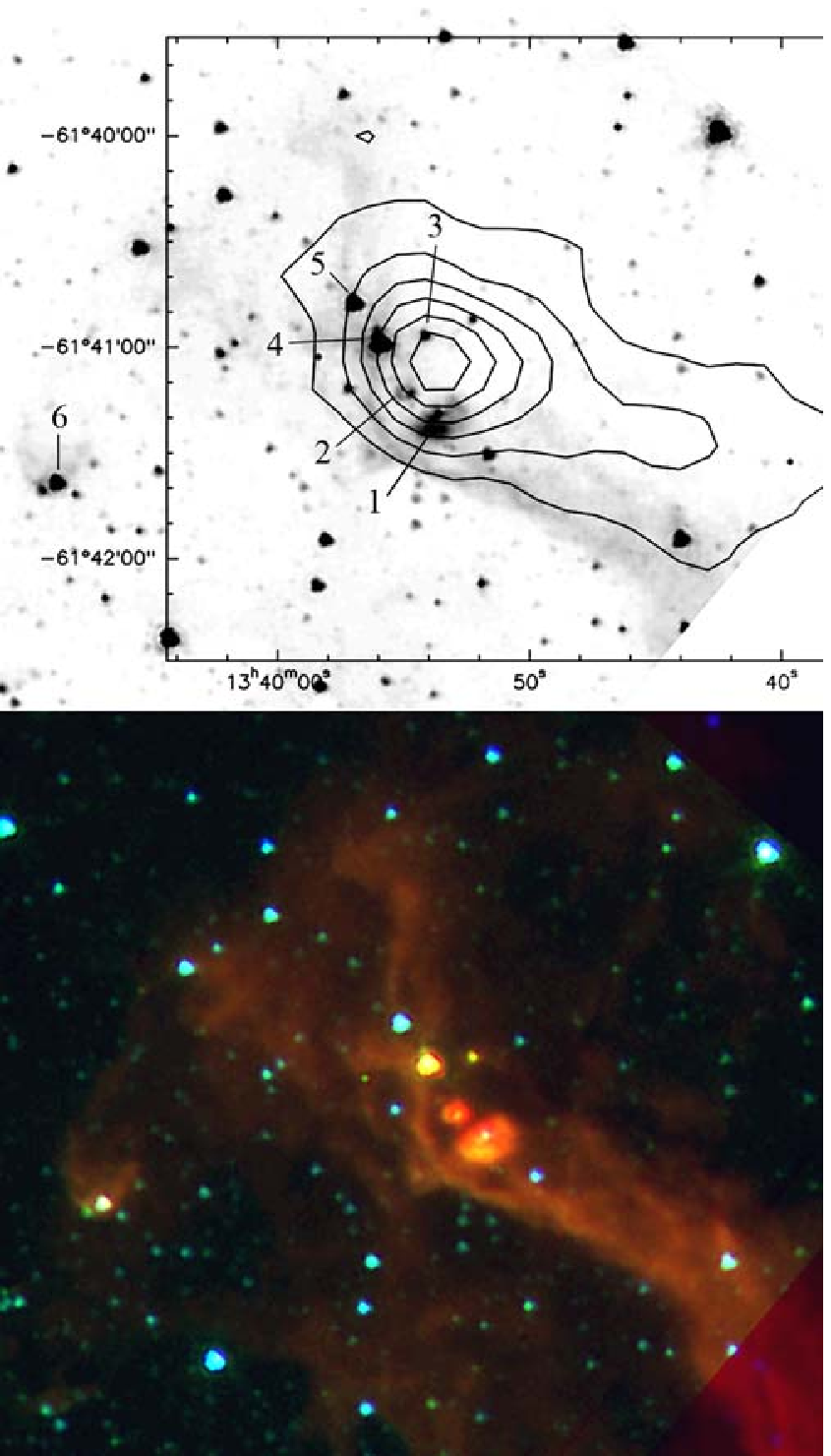}
  \caption{As Fig.~\ref{cond2smmf} but for condensation 4  }
  \label{cond4smmf}
\end{figure}

Figs.~\ref{cond2smmf}, \ref{cond3smmf}, and \ref{cond4smmf} (top)
present the 1.2-mm emission as contours superimposed on the Spitzer
GLIMPSE 3.6\,$\mu$m frame for condensations 2, 3 and 4,
respectively. Several of the objects discussed in the text are
identified according to their numbers in Table~\ref{spitph}.
Figs.~\ref{cond2smmf}, \ref{cond3smmf}, and \ref{cond4smmf} (bottom)
present a colour composite image displaying the 2MASS $K_{S}$ frame
in blue, the GLIMPSE 3.6\,$\mu$m frame in green and the GLIMPSE
8\,$\mu$m frame in red, for condensations 2, 3 and 4, respectively.
Note that the 3.6 and 8\,$\mu$m images are very similar, both
filters including PAH emission bands (at 3.3 and 8.6\,$\mu$m).
However, the continuum emission from normal stars is only visible in
the 3.6\,$\mu$m image and not at 8\,$\mu$m, hence a dominance of the
$K_{S}$ (blue) and 3.6\,$\mu$m (green) emissions from normal stars
in Figs.~\ref{cond2smmf}, \ref{cond3smmf}, and \ref{cond4smmf}
(bottom).

\begin{figure}[tb]
  \includegraphics[angle=270,width=85mm ]{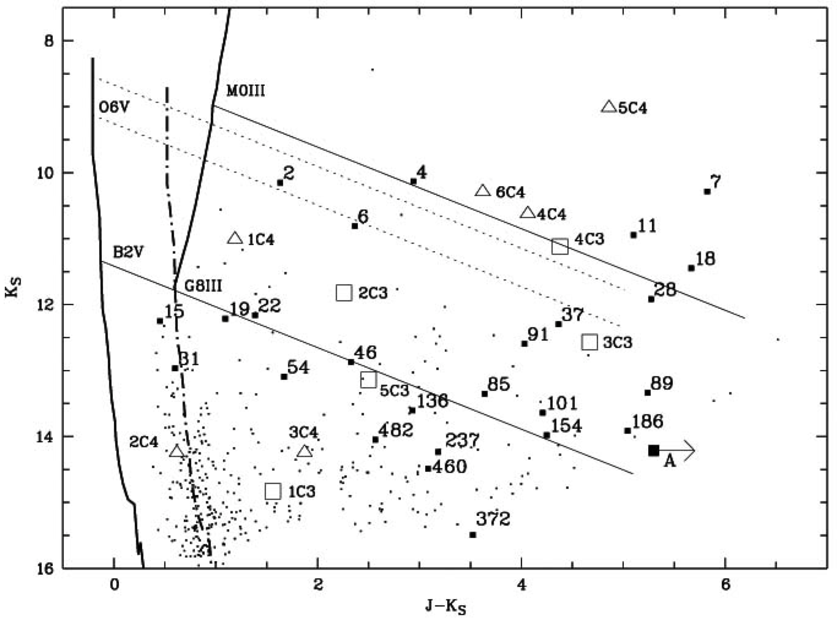}
  \caption{Magnitude-colour diagram for the sources observed towards condensations 2, 3 and 4. The thick solid curves show the main sequence (class V)
  and the sequence of giants (class III) for a distance of 4.3~kpc and no extinction. The thick dash-dotted curve shows the main sequence with
  a foreground visual extinction of 4.3~mag (see text).
  The two solid lines are the reddening lines for 30~mag of visual extinction for a
  B2V star
  and an M0III star. The thin dotted lines indicate the extreme values
  for the absolute magnitude of an O6V star with reddening
  lines corresponding to 30~mag of visual extinction }
  \label{magc}
\end{figure}

\begin{figure}[tb]
  \includegraphics[angle=270,width=85mm ]{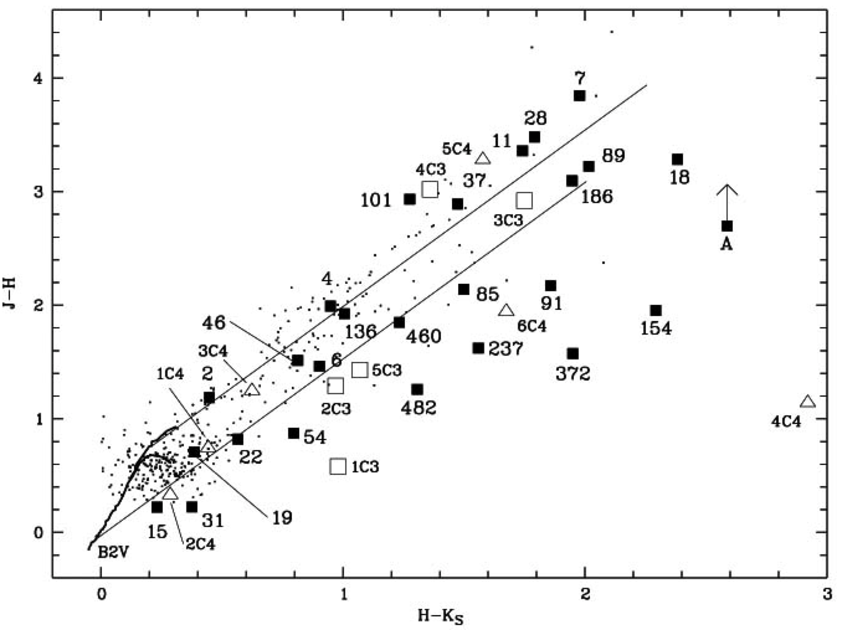}
  \caption{Colour-colour diagram for sources observed towards condensations 2, 3 and 4. The two curves show the main sequence (class V)
  and the sequence of giants (class III) for a distance of 4.3~kpc and no extinction. The parallel lines are the reddening lines
  for 30~mag of visual extinction for a B2V star (lower) and an M0III star (upper)  }
  \label{cc}
\end{figure}

\begin{figure}[tb]
  \includegraphics[angle=-90,width=85mm ]{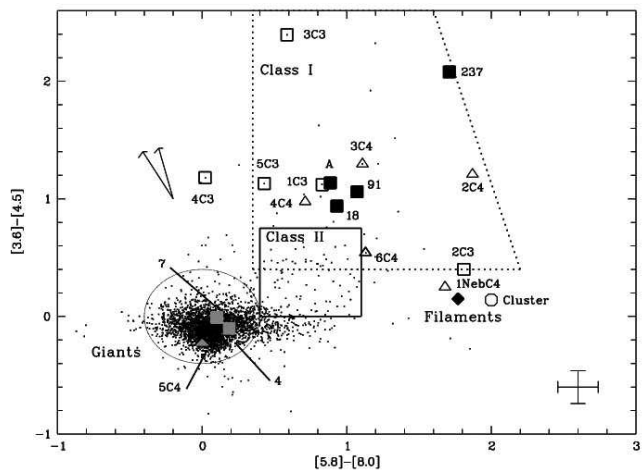}
  \caption{GLIMPSE colour-colour diagram, [3.6]$-$[4.5] versus [5.8]$-$[8.0], for
  the sources observed towards the most massive condensations
  detected at 1.2-mm (nos 2, 3 and 4) and the sources observed in the RCW~79 field (small black dots), extracted from the GLIMPSE
  PSC. The sources observed towards the compact \HII\ region are
  identified by their numbers in Table~\ref{spitph} and shown as
  filled squares. The sources observed towards condensation 3 are
  shown as empty squares and identified by their numbers in
  Table~\ref{spitph} plus C3 (i.e. 1C3 for object 1 of condensation
  3). Same for sources observed towards condensation 4, shown as
  empty triangles.
  The locations of Class~I and II YSOs and red giants are from Allen et al. (2004) and Whitney et al.
  (2004). Extinction vectors are shown for $A_V$=30~mag, taken from
  Allen et al. (2004).
  The stars discussed in the text are identified. Typical error bars are shown}
  \label{spitcc}
\end{figure}

Table~\ref{spitph} gives the position and photometry from
1.25\,$\mu$m to 8\,$\mu$m of the sources discussed in the following.
Column~1 gives their identification numbers, according to
Figs.~\ref{cond2smmf} (see also Fig.~\ref{zonesk}), \ref{cond3smmf},
and \ref{cond4smmf}, for condensations 2, 3 and 4, respectively.
Columns~2 and 3 give their coordinates according to the GLIMPSE
point source catalogue (PSC, http://www.astro.wisc.edu/sirtf/),
except for sources in condensation 2 for which the ESO-NTT positions
- more accurate - have been taken. For the star of source 1 in
condensation 4 (1~star in Table~\ref{spitph}) we have taken
the 2MASS position. Columns 4 to 6
gives their $J$, $H$ and $K_{S}$ magnitudes, from the ESO-NTT
observations for sources near condensation 2, and from the 2MASS PSC
(http://tdc-www.harvard.edu/software/catalogs/tmpsc.html) for
sources near condensations 3 and 4. Columns 7 to 10 gives their
[3.6], [4.5], [5.8] and [8.0] magnitudes from the GLIMPSE PSC. When
not available from the PSC, GLIMPSE magnitudes have been measured
(aperture photometry) using the Basic Calibrated Data frames. Those
measurements are indicated with asterisks in Table~\ref{spitph}.
% Table~3
\begin{table*}
%{\tiny{
\caption{Magnitudes of sources associated with the 1.2-millimetre
condensations 2, 3 and 4}
\begin{tabular}{lllrrrrrrrl}
\hline \hline
Object & \multicolumn{1}{c}{$\alpha_{2000}$} & \multicolumn{1}{c}{$\delta_{2000}$} & $J$ & $H$ & $K_{S}$ & [3.6] & [4.5] & [5.8] & [8.0] & Comments\\
  \hline
\multicolumn{11}{c}{\bf Condensation 2} \\
      &                &                &         &         &              &    &   &   \\
  \multicolumn{11}{l}{Cluster} \\
 &                &                &         &         &              & & & &   \\
    6   & 13$^{\rm h}$ 40$^{\rm m}$ 53$\fs$8 &  $-$61$\degr$ 45$\arcmin$ 46$\arcsec$ & 13.17 & 11.71 & 10.81 & & & & & MS star, $A_{V}$=15~mag   \\
    11 &  13$^{\rm h}$ 40$^{\rm m}$ 53$\fs$1 &  $-$61$\degr$ 46$\arcmin$ 11$\arcsec$ & 16.05 & 12.69 & 10.94 & 9.75 & 9.77  & 9.41 &  & Giant, $A_{V}$=23~mag        \\
    15 & 13$^{\rm h}$ 40$^{\rm m}$ 54$\fs$3 &  $-$61$\degr$ 45$\arcmin$ 47$\arcsec$ & 12.70 & 12.48 & 12.25  & & & & &  MS star, $A_{V}$=3~mag  \\
    19 & 13$^{\rm h}$ 40$^{\rm m}$ 53$\fs$3 &  $-$61$\degr$ 45$\arcmin$ 53$\arcsec$ & 13.30 & 12.60 & 12.21 & & & & & MS star, $A_{V}$=7~mag  \\
    22 & 13$^{\rm h}$ 40$^{\rm m}$ 52$\fs$5 &  $-$61$\degr$ 45$\arcmin$ 52$\arcsec$ & 13.55 & 12.73 & 12.16 & & & & & MS star, $A_{V}$=9~mag             \\
    46  & 13$^{\rm h}$ 40$^{\rm m}$ 53$\fs$5 &  $-$61$\degr$ 45$\arcmin$ 47$\arcsec$ & 15.19 & 13.68 & 12.87 & & & & &MS star, $A_{V}$=14.5~mag \\
    85  & 13$^{\rm h}$ 40$^{\rm m}$ 54$\fs$9 &  $-$61$\degr$ 45$\arcmin$ 48$\arcsec$ & 16.99 & 14.85 & 13.35 & & & & & MS star, $A_{V}$=22~mag    \\
   136  & 13$^{\rm h}$ 40$^{\rm m}$ 53$\fs$1 &  $-$61$\degr$ 45$\arcmin$ 51$\arcsec$ & 16.53 & 14.61 & 13.60 & & & & & MS star, $A_{V}$=18~mag         \\
    154 & 13$^{\rm h}$ 40$^{\rm m}$ 54$\fs$0 &  $-$61$\degr$ 45$\arcmin$ 49$\arcsec$ & 18.22 & 16.27&  13.98 & & & & &  Near-IR excess            \\
    372 & 13$^{\rm h}$ 40$^{\rm m}$ 54$\fs$4 &  $-$61$\degr$ 45$\arcmin$ 50$\arcsec$ & 19.02 & 17.44 & 15.49 & & & & &  Near-IR excess            \\
   &                &                &         &         &              & & & & &   \\
    Whole cluster & 13$^{\rm h}$ 40$^{\rm m}$ 53$\fs$4 &  $-$61$\degr$ 45$\arcmin$ 53$\arcsec$ &  & & & 7.35$^{*}$ & 7.21$^{*}$ & 4.74$^{*}$ & 2.74$^{*}$
  & \\
 &                &                &         &         &              &    &   & &   \\
   237  & 13$^{\rm h}$ 41$^{\rm m}$ 02$\fs$3 &  $-$61$\degr$ 44$\arcmin$ 16$\arcsec$ & 17.41  & 15.79 & 14.23 & 10.64 & 8.56 & 7.79
   & 6.08 & Near-IR excess  \\
   &                &                &         &         &              &    &   & &   \\
  \multicolumn{11}{l}{Filament} \\
&                &                &         &         &              & & & & &   \\
    31  & 13$^{\rm h}$ 40$^{\rm m}$ 56$\fs$6 &  -61$\degr$ 45$\arcmin$ 39$\arcsec$ & 13.56 & 13.34 & 12.96 & 12.49 & 12.60 & & &            \\
    54 & 13$^{\rm h}$ 40$^{\rm m}$ 56$\fs$8 &  -61$\degr$ 45$\arcmin$ 45$\arcsec$ & 14.76 & 13.89 & 13.09  &  & & & &           \\
91 & 13$^{\rm h}$ 40$^{\rm m}$ 57$\fs$6 &  -61$\degr$ 45$\arcmin$ 44$\arcsec$ &  16.62 & 14.45 & 12.59  & 8.66$^{*}$ & 7.60$^{*}$ & 7.25 & 6.18 & Maser position      \\
460 & 13$^{\rm h}$ 40$^{\rm m}$ 57$\fs$9 &  -61$\degr$ 45$\arcmin$ 42$\arcsec$ & 17.57 & 15.72 & 14.49 & & & & &            \\
482 & 13$^{\rm h}$ 40$^{\rm m}$ 58$\fs$2 &  -61$\degr$ 45$\arcmin$ 45$\arcsec$ & 16.61 & 15.35 & 14.05 &  & & & &           \\
 &                &                &         &         &              &  & & & & \\
  \multicolumn{11}{l}{Red objects} \\
&                &                &         &         &              & & & & &  \\
18 & 13$^{\rm h}$ 40$^{\rm m}$ 58$\fs$3 &  -61$\degr$ 46$\arcmin$ 50$\arcsec$ & 17.11 & 13.83 & 11.45 & 8.47 & 7.54 & 6.79 & 5.85 & Near-IR excess         \\
89 & 13$^{\rm h}$ 40$^{\rm m}$ 58$\fs$7 &  -61$\degr$ 46$\arcmin$ 52$\arcsec$ & 18.58 & 15.35 & 13.34  &  10.83  & 10.71 & & & Giant?        \\
186 & 13$^{\rm h}$ 40$^{\rm m}$ 58$\fs$5 &  -61$\degr$ 46$\arcmin$ 48$\arcsec$ & 18.96 & 15.86 & 13.91 &   & & & & Giant?          \\
A &  13$^{\rm h}$ 41$^{\rm m}$ 00$\fs$0 &  -61$\degr$ 46$\arcmin$ 58$\arcsec$ & $\geq$ 19.5 & 16.80 & 14.21 & 10.95 & 9.81 & 8.72 & 7.84 & Not detected in $J$     \\
101 & 13$^{\rm h}$ 40$^{\rm m}$ 59$\fs$0 &  -61$\degr$ 46$\arcmin$ 50$\arcsec$ & 17.85 & 14.91 & 13.64 & & & & & Giant           \\
 &                &                &         &         &              &  & & & & \\
  \multicolumn{11}{l}{Giants} \\
&                &                &         &         &              &  & & & & \\
2 & 13$^{\rm h}$ 41$^{\rm m}$ 04$\fs$5 &  -61$\degr$ 44$\arcmin$ 41$\arcsec$ & 11.78 & 10.60 & 10.15 & 9.82   & 10.03 & 10.03 & &          \\
4 & 13$^{\rm h}$ 41$^{\rm m}$ 02$\fs$6 &  -61$\degr$ 44$\arcmin$ 45$\arcsec$ & 13.07 & 11.08 & 10.13 & 9.50   & 9.60 & 9.41 & 9.71 &          \\
7 & 13$^{\rm h}$ 41$^{\rm m}$ 05$\fs$2 &  -61$\degr$ 44$\arcmin$ 46$\arcsec$ & 16.11 & 12.262 & 10.28 & 8.87   & 8.88 & 8.59 & 8.49 &          \\
28 &  13$^{\rm h}$ 41$^{\rm m}$ 05$\fs$3 &  -61$\degr$ 44$\arcmin$ 35$\arcsec$ & 17.19 & 13.71 &  11.91 & 10.60  & 10.67 & & &           \\
37 & 13$^{\rm h}$ 41$^{\rm m}$ 05$\fs$4 &  -61$\degr$ 44$\arcmin$ 59$\arcsec$ & 16.66 & 13.77 & 12.30 & 11.30  & 11.23 & 11.54 & &           \\
&  &  &  &  &  &  &  &  &  & \\
\multicolumn{11}{c}{\bf Condensation 3} \\
 &  &  &  &  &  &  &  &  &  \\
 1 & 13$^{\rm h}$~40$^{\rm m}$~38$\fs3$ & $-$61$\degr$~47$\arcmin$~00$\arcsec$ &
 16.39 & 15.81  & 14.83 & 9.73 & 8.61 & 7.62 & 6.79 & Near-IR excess \\
 2        & 13$^{\rm h}$~40$^{\rm m}$~32$\fs$65 & $-$61$\degr$~47$\arcmin$~20$\arcsec$ &
  14.08 & 12.79 & 11.82 & 9.43 & 9.03 & 6.76 & 4.95 & Near-IR excess  \\
 3        & 13$^{\rm h}$~40$^{\rm m}$~26$\fs$6 & $-$61$\degr$~47$\arcmin$~56$\arcsec$ &
  17.24 & 14.32 & 12.57 & 9.55 & 7.15 & 5.89 & 5.29 & \\
 4        & 13$^{\rm h}$~40$^{\rm m}$~25$\fs$2 & $-$61$\degr$~47$\arcmin$~48$\arcsec$ &
  15.50 & 12.48 & 11.12 &  9.03 & 7.85 & 7.05 & 7.03 & Giant \\
 5        & 13$^{\rm h}$~40$^{\rm m}$~27$\fs$5 & $-$61$\degr$~47$\arcmin$~50$\arcsec$ &
  15.64 & 14.21 & 13.14 & 9.80 & 8.67 & 7.84 & 7.41 & Near-IR excess  \\
&  &  &  &  &  &  &  &  &  & \\
\multicolumn{11}{c}{\bf Condensation 4} \\
 &  &  &  &  &  &  &  &  &  & \\
 1~nebula & 13$^{\rm h}$~39$^{\rm m}$~53$\fs6$ & $-$61$\degr$~41$\arcmin$~21$\arcsec$ &
  &  & &  8.19$^{*}$ & 7.94$^{*}$ & 5.09$^{*}$ & 3.41$^{*}$ & \\
 1~star   & 13$^{\rm h}$~39$^{\rm m}$~53$\fs$68 & $-$61$\degr$~41$\arcmin$~19$\farcs$2 &
  12.19 & 11.45 & 11.01 & & & & & \\
 2        & 13$^{\rm h}$~39$^{\rm m}$~54\fs7 & $-$61\degr~41\arcmin~13\arcsec &
  14.86 & 14.53 & 14.25 & 10.90 & 9.69 & 7.15 & 5.28  &\\
 3        & 13$^{\rm h}$~39$^{\rm m}$~54\fs2 & $-$61\degr~40\arcmin~57\arcsec &
  16.11 & 14.86 & 14.24 & 10.32 & 9.02 & 7.83 & 6.72  & \\
 4        & 13$^{\rm h}$~39$^{\rm m}$~55\fs9 & $-$61\degr~40\arcmin~58\arcsec &
  14.68 & 13.54 & 10.62 &  7.19 & 6.22$^{*}$ & 5.40 & 4.69 &  Near-IR excess \\
 5        & 13$^{\rm h}$~39$^{\rm m}$~57\fs1 & $-$61\degr~40\arcmin~47\arcsec &
  13.87 & 10.59 & 9.01 & 7.80 & 8.02 & 7.62 & 7.62 & Giant \\
 6        & 13$^{\rm h}$~40$^{\rm m}$~08\fs9 & $-$61\degr~41\arcmin~38\arcsec &
  13.91 & 11.96 & 10.29 & 8.72 & 8.18 & 7.51 & 6.38 & Near-IR excess  \\
 \hline
\end{tabular}
\label{spitph}
\\

{$^{\footnotesize *}$ Magnitude obtained by aperture photometry (see text)}   \\
\\
\\

\end{table*}
We use both near- and mid-IR data to discuss the nature of the
sources observed towards the condensations. We present the $K_{S}$
versus $J-K_{S}$, $J-H$ versus $H-K_{S}$ and the [3.6]$-$[4.5]
versus the [5.8]$-$[8.0] diagrams of the stars observed towards
condensations 2, 3 and 4 in Figs.~\ref{magc}, \ref{cc} and
\ref{spitcc}, respectively. The sources are identified in
Figs.~\ref{magc}, \ref{cc} and \ref{spitcc} as their number in
Table~\ref{spitph} at which we attached C3 or C4 for labeling the
objects of condensation 3 and 4, respectively. For the main sequence
in Fig.~\ref{magc} and \ref{cc}, we have used the absolute
magnitudes $M_{V}$ of stars from O3 to O9 of Martins et
al.~(\cite{mar05}) and from Schmidt-Kaler (\cite{sch82}) for later
type stars. The colours for main sequence stars and the colours and
absolute magnitudes for giants are from Koornneef (\cite{koo83}).
Note that large differences exist in the estimated absolute visual
magnitudes of massive stars, depending on the authors. As an example
we display in Fig.~\ref{magc} the uncertainties for an O6V star. The
absolute magnitude of such a star is $-$5.5 according to
Schmidt-Kaler, $-$5.11 according to Vacca et al.~(\cite{vac96}) and
$-$4.92 according to Martins et al.~(\cite{mar05}). Recent results
on ISO-SWS data dealing with ionization diagnostics of compact \HII\
regions (Morisset et al.~\cite{mor04}) indicate that the model used
by Martins et al. gives better agreement with the observations.
Therefore we adopt this calibration in the following. Typical errors
are 0.04 for the $K_{S}$ magnitudes and 0.07 for the $J-H$ and
$H-K_{S}$ colours. The reddening law is from Mathis (\cite{mat90})
for $R_{V}$=3.1. In both diagrams we have drawn the interstellar
reddening line originating from a B2V star and from a MOIII star,
for $A_{V}$=30~mag. In Fig~\ref{magc}, we also plot the main
sequence shifted by a visual extinction of $A_{V}$=4.3~mag using the
standard value of $A_{V}$=1~mag kpc$^{-1}$ for the foreground
interstellar reddening.

We are interested in detecting massive young stellar objects and
candidates ionizing stars. Therefore we identify, in
Fig.~\ref{magc}, all the sources that have a $K_{S}$ magnitude
brighter than that corresponding to a B2V star. However, giants also
fall in this group and the magnitude-colour diagram alone does not
allow one to separate those stars from the others. Therefore we use
the colour-colour diagrams (Figs.~\ref{cc} and \ref{spitcc}), which
clearly separates giants from other sources.

Fig.~\ref{spitcc} presents the GLIMPSE colour-colour diagram for the
sources of Table~\ref{spitph} and for all the sources observed in
the direction of the RCW~79 field, extracted from the GLIMPSE PSC.
The locations of Class~I sources (SED dominated by emission from an
envelope), Class~II sources (SED dominated by emission from a disk),
and giants are taken from Allen et al.~(\cite{all04}) and Whitney et
al.~(\cite{whi04}). The $[3.6]-[4.5]$ versus $[5.8]-[8.0]$ diagram
is a useful tool for identifying young sources in different
evolutionary stages. The geometry of the source and the temperature of
the central object deeply influence its location in this diagram
(see Whitney et al.~\cite{whi04}). Two extinction vectors are shown
in Fig.~\ref{spitcc} for $A_{V}$=30~mag, taken from Allen et al.~(\cite{all04}),
using the two extremes of six vectors calculated by Megeath et al.~(\cite{meg04}).
Extinction tends to move sources to the upper left (see Fig.~\ref{spitcc}).
We have also indicated in Fig.~\ref{spitcc} the colours of the filaments (filled lozenge)
measured in the direction of the PDR surrounding the ionized gas,
and probably dominated by the PAH emission bands in the 3.6\,$\mu$m,
5.8\,$\mu$m and 8.0\,$\mu$m filters. These colours are
$[3.6]-[4.5]\simeq$0.1 and $[5.8]-[8.0]\simeq$1.8. \\

\noindent {\bf{Condensation 2}}

\begin{figure}[tb]
  \includegraphics[angle=0,width=85mm]{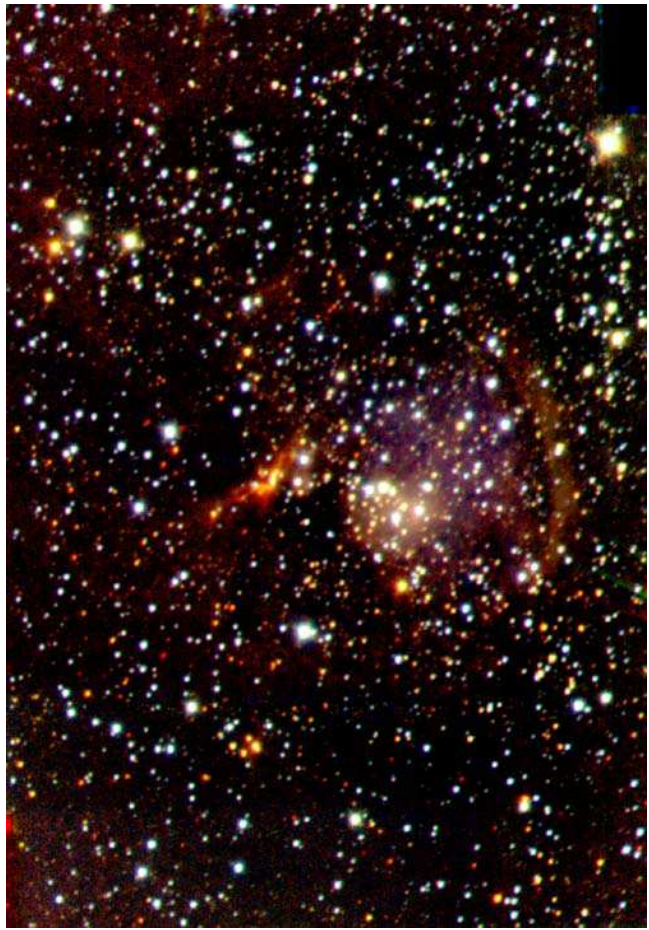}
  \caption{$J$ (blue), $H$ (green) and $K_{S}$ (red) composite colour image of the cluster ionizing the compact
  \HII\ region (NTT observations). The field size is $2\farcm66\times3\farcm85$. North is up, east is
  left}
  \label{jhk}
\end{figure}

\begin{figure}[tb]
  \includegraphics[angle=0,width=85mm]{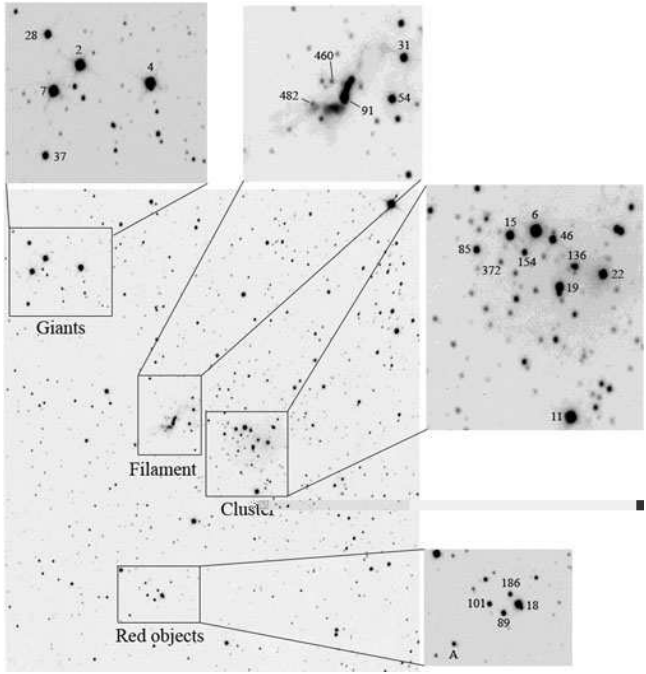}
  \caption{NTT $K_S$ image centred on the compact \HII\ region. The regions and stars discussed in the text are identified}
  \label{zonesk}
\end{figure}

Many red objects are observed in the direction of condensation 2
(Figs.~\ref{cond2smmf} and \ref{jhk}) but {\it none are observed at
the condensation's millimetre emission peak}.

Figure~\ref{jhk} shows a $JHK_{S}$ colour composite image of the
compact \HII\ region observed on the border of RCW~79. This colour
image outlines four main zones that are identified in
Fig.~\ref{zonesk} and discussed below.

{\underline{\it{The stellar cluster exciting the compact \HII\
region}}} As shown in Fig.~\ref{zonesk}, stars 6, 15, 19, 22, 46,
85, 136, 154 and 372 are observed in the direction of the central
cluster. The nature and visual extinction of those sources, derived
from the magnitude-colour and colour-colour diagrams, are given in
Table~\ref{spitph}. Stars 6, 15, 19, 22, 46, 85 and 136 are probably
main-sequence stars affected by a visual extinction in the range 3
-- 22~mag, possibly indicating large small-scale variations of the
visual extinction. Note that massive stars have no pre-main sequence
phase and are already on the main-sequence while still accreting matter. Inside
the cluster, stars 6 and 22 are bright at all wavelengths. Star 6 is
the dominant exciting star of the compact \HII\ region. According to
the absolute calibration of Martins et al.~(\cite{mar05}) its $K$
magnitude points to a spectral type between O6 and O7
(Fig.~\ref{magc}), hence earlier than what is derived from the radio
observations (Sect.~\ref{rad}). However, part of the stellar
radiation may escape and not be used to ionize the gas, leading to
an underestimate of the spectral type using radio observations.
Near-IR spectroscopy is needed to discuss the spectral type of young
massive stars (Repolust et al.\ \cite{rep05}). Star 22 seems to be
associated with diffuse emission on the 4.5\,$\mu$m and 8\,$\mu$m
frames of the GLIMPSE survey. (However the GLIMPSE resolution does
not allow determination of the magnitudes of the individual stars in
the cluster). Stars~154 and 372 show a near-IR excess. The
colour-colour diagram of Fig.~\ref{cc} shows that star 11 is a
giant. This is confirmed by its $[3.6]-[4.5]$ colour, and by its
non-detection at 8\,$\mu$m.

A faint red arc is observed in the $K_{S}$ frame, north-west of the
cluster centre (Fig.~\ref{jhk}). This arc outlines the limits of the compact \HII\
region. The origin of this red emission may be fluorescent H$_2$
emission from the hot PDR and included in the $K_{S}$ band at
2.12\,$\mu$m. H$_2$ at 2.12\,$\mu$m and dust emission are known to
coincide spatially in Galactic PDRs due to a common excitation
origin from UV photons and to the efficient formation of H$_{2}$
molecules on the surface of grains (Habart et al. \cite{hab03},
Zavagno \& Ducci \cite{zav01}). The complete emission shell that
outlines the ionization front is not observed in $K_{S}$ but clearly
appears at 8\,$\mu$m in the GLIMPSE frame (Fig.~\ref{cond2smmf}
bottom). This indicates that the zone surrounding the central
cluster is more affected by extinction in the south and east.

 {\underline{{\it{The filament}}}}
%---------------------------
A red filament is observed 57$\arcsec$ (1.2~pc) north-east of the
central cluster (Figs.~\ref{jhk} and \ref{zonesk}). The emission
peak of this filament corresponds to object~91, which shows a
near-IR excess (Fig.~\ref{cc}). Maser sources are observed in the
exact direction of object 91 (see Sect.~2). Furthermore
the near-IR colours of object 91 are in good agreement with the ones
found in high-mass star forming regions, at the positions of maser
emissions (Goedhart et al.~\cite{goe02}). Object 91 is a bright
Class~I object (Figs.~\ref{cond2smmf} and \ref{spitcc}). All this
shows that massive-star formation is presently taking place in this
region.

Other objects observed towards this region (31, 54, 482) show a
near-IR excess, reinforcing the idea of an active and recent star
forming region. Object 54 seems to be associated with a small
nebulosity, conspicuous on the 8\,$\mu$m GLIMPSE image
(Figs.~\ref{cond2smmf} bottom).

{{\underline{\it{The red objects group}}}}
%------------------------------------
A group of four very red objects (18, 89, 101, 186) is observed
south-east of the central cluster (see Fig.~\ref{zonesk}). An
additional isolated red source (object A) is observed nearby. This
zone is located at the south border of condensation 2
(Fig.~\ref{cond2smmf} top). As seen in Fig.~\ref{cc} object 101 is a
reddened giant. Sources 18 and A have a near-IR excess. Source A is
the reddest object of the field, with $H-K_{S}=2.59$ (it is not
detected in $J$ which indicates that $J\geq$ 19.5 and therefore $J-H
\geq$ 2.7). Sources 18 and A brighten at longer wavelengths, as
shown by the GLIMPSE data (Fig.~\ref{cond2smmf}). Fig.~\ref{spitcc}
shows that sources A and 18 are Class I objects with rising SEDs.

The nature of objects 89 and 186 is not clear. They are brighter
than B2V stars, but they have no associated \HII\ regions; according
to Fig.~\ref{cc} they do not present a near-IR excess. They are very
reddened, thus probably are not foreground objects. Star 89 has the
$[3.6]-[4.5]$ colour of a giant, and both sources disappear at
longer wavelengths; we propose that both are giants, even if it is not
clear from Fig.~\ref{cc}.

Many other red sources are observed in the near IR. Those sources
are probably located behind the filamentary dust structures observed
in the mid-IR that outline the PDR associated with the compact
\HII\ region (see Fig.~\ref{cond2smmf}). These objects are too faint
to be detected on the GLIMPSE images and are not, therefore,
discussed here.

{{\underline{\it{Reddened giants}}}}
%---------------------------------------------
A group of reddened giants is observed on the upper north-eastern
part of the field (Fig.~\ref{zonesk}). This zone contains very
luminous objects (2, 4, 7, 28, 37) as shown by Fig.~\ref{magc}. The
colour-colour diagram of Fig.~\ref{cc} reveals that all those
sources are reddened giants affected by a visual extinction in the
range 5--27~mag. Those reddened sources are visible in the GLIMPSE
images and possess specific colours (Fig.~\ref{spitcc}) that make
them easy to identify as giants (Indebetouw et al.\ \cite{ind05}).

Object 237 is particularly interesting as a bright Class~I source
that appears isolated on the GLIMPSE images (Fig.~\ref{cond2smmf}).
It is a faint $K_{S}$ source with a near-IR excess (Fig.~\ref{cc})
and a strongly rising SED. This source is probably a massive young
stellar object.
\\

 \noindent {\bf{Condensation 3}}
%-----------------------------

As seen in Fig.~\ref{cond3smmf}, several luminous red objects are
observed in the direction of condensation~3. Objects 1, 3 and 5
(1C3, 3C3 and 5C3, respectively) are Class I sources
(Fig.~\ref{spitcc}). Object 1 is relatively faint in $K_{S}$ and has
a near-IR excess (Fig.~\ref{cc}). Object 3 is highly reddened and is
probably a high mass object. Object 5 has a near-IR excess
(Fig.~\ref{cc}) confirming its nature as a young star. Object 2 has
GLIMPSE colours very similar to those of the filaments. It may be a
B star surrounded by a small PDR. From Fig.~\ref{spitcc} the nature
of object 4 is not clear, possibly due to the proximity of another
star. However, the near-IR data indicate that this source may be a
giant (Figs.~\ref{cc}, \ref{magc}). \\

\noindent {\bf{Condensation 4}}
%-----------------------------

As seen in Fig.~\ref{cond4smmf}, several luminous red objects lie
along the border of condensation 4, close to the ionized gas, but
{\it none are observed at the condensation's millimetre emission
peak}. These objects are of different nature. Object 1 (1C4) is a
small nebulosity of about 0.3~pc\,$\times$\,0.4~pc around a central
star. Its GLIMPSE colours are very similar to those of the
filaments. This extended emission probably originates from a PDR
region associated with star 1, a B star with a visual extinction of
7~mag. The near-IR data indicate a massive main-sequence source.
Object 2 (2C4) is also a small nebulosity with the colours of the
filaments. This source is relatively faint in the $K_{S}$ band and
may have a near-IR excess if it is a low mass object. Objects 3
(3C3)and 4 (4C4) are Class~I objects (Fig.~\ref{spitcc}). Object 4
is particularly interesting as it is a high luminosity source, one
of the brightest objects seen by GLIMPSE in the RCW~79 field, apart
from giants. This object has a large near-IR excess ((Fig.~\ref{cc})
and is probably a young massive star. Object 5 (5C4) is a red giant
with a visual extinction of 20~mag. This is confirmed by its
position in Figs.~\ref{magc}, \ref{cc} and \ref{spitcc}, lying in
the giant region where numerous stars are observed. Object 6 (6C4)
is situated at the head of a bright rim surrounding condensation 6
(Fig.~\ref{cond4smmf}). It is a Class~I or Class~II object
(Fig.~\ref{spitcc}). This source is bright in $K_{S}$
(Fig.~\ref{magc}) and has a near-IR excess (Fig.~\ref{cc}).

No stars are observed at the millimetre peak of the condensations.
This indicates either that no stars are there at all, or that
earlier phases of star formation have taken place but are not
detected in the mid IR due to the low temperature of the sources and
the high extinction. Typical spectral energy distribution of Class~0
sources shows mid-IR fluxes below the GLIMPSE detection limits.
Deeper mid-IR observations are needed to address the question of
star formation at condensations millimetre peak.

Fig.~\ref{spitcc} shows that twelve Class~I sources, associated with
the three most massive millimetre condensations, are observed at the
periphery of RCW~79. Some of those sources are very bright and have
near-IR excess and are thus possibly massive young stellar objects.
This result indicates that we are observing a case of relatively
recent massive-star formation at the borders of RCW~79.

\subsection{The velocity field of the ionized gas \label{vha}}
\begin{figure}[tb]
  \includegraphics[angle=0,width=85mm]{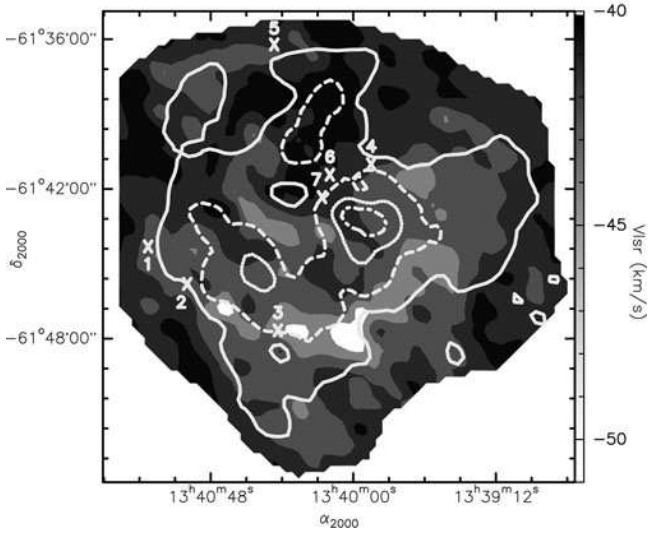}
  \caption{Grey-scale image of the H$\alpha$ velocity field of RCW~79. The map is overlaid with
H$\alpha$ intensity contours (in arbitrary units). The peak
positions of the 1.2-mm condensations are indicated (white cross)}
  \label{havel}
\end{figure}

Figure \ref{havel} presents the H$\alpha$ velocity field of RCW~79.
The zones of brightest H$\alpha$ emission have a mean velocity of
$-$44~km~s$^{-1}$. An annular structure of more diffuse emission,
with a velocity $\sim$ $-$50~km~s$^{-1}$, limits the \HII\ region;
especially, at the southern border of RCW~79, a clear arc-like
structure is observed at $-$51~km~s$^{-1}$. The H$\alpha$ emission
observed at 13$^{\rm{h}}$40$^{\rm{m}}$10$\fs$3,
$-$61$\degr$38$\arcmin$35$\arcsec$, in the direction of the ``hole''
observed in the dust ring, has a velocity of $-$40~km~s$^{-1}$.
Thus, the shape of the dust ring surrounding the ionized region,
with its conspicuous hole, and the velocity field showing a flow of
ionized material at a few km~s$^{-1}$, suggest that the molecular
environment of RCW~79 is broken in the north-west, and that the
ionized gas is flowing away from the centre of the \HII\ region
through this hole. This is coherent with the observed location of
the most massive condensations to the south and west, limiting the
ionized zone more efficiently in this direction. The presence of a
molecular cloud to the south of RCW~79 (Saito et al.~\cite{sai01})
corroborates this interpretation.
The observed $-$51~km~s$^{-1}$ to $-$40~km~s$^{-1}$ H$\alpha$
velocity range is in good agreement with the velocities measured
from the H109$\alpha$ radio recombination line by Wilson et
al.~(\cite{wil70}, $-$46.4 and $-$51.8~km~s$^{-1}$, 4$\arcmin$
HPBW). Similar velocities are measured for the associated molecular
gas, from the CS (2 -- 1) emission line by Bronfman et
al.~(\cite{bro96}, $-$48~km~s$^{-1}$, 50$\arcsec$ HPBW), and for the
methanol maser observed in the direction of IRAS~13374$-$6130 by
Caswell~(\cite{cas04}, $-$51~km~s$^{-1}$).

CGPMC report a lack of H\,{\sc{i}} emission, obtained from the
Southern Galactic Plane Survey (SGPS, McClure-Griffiths et
al.~\cite{mcc05}) observed between $-$26.4 and $-$28.9~km s$^{-1}$
(see their fig.~13). The shape of this absorption strongly suggests
that it is associated with the \HII\ region RCW~79. We checked the
H\,{\sc{i}} data cube and observed an H\,{\sc{i}} absorption at
$-$25.6~km s$^{-1}$ that has the same structures as the ones seen in the
H$\alpha$ emission, especially the northern H$\alpha$ flow. However
the H\,{\sc{i}} velocity does not correspond to that observed for
the ionized gas. A problem of H\,{\sc{i}} velocity calibration may
occur in this case.

\section{Comparison with the model of Whitworth et al.}
%----------------------------------------------------------
Whitworth et al.~(\cite{whi94}) have developed an analytical model
to describe the fragmentation of the shocked dense layer surrounding
an expanding \HII\ region. In particular, these authors predict the
time at which the fragmentation occurs, the size of the \HII\ region
at that moment, the column density of the layer, the masses of the
fragments and their separations along the layer. The adjustable
parameters of the model are $\dot{\cal N}_{\rm{\tiny{Lyc}}}$,  the
number of Lyman continuum photons emitted per second by the exciting
star, the density $n_{\tiny{{0}}}$ of the surrounding homogeneous
infinite medium into which the \HII\ region expands, and
$a_{\rm{\tiny{S}}}$, the isothermal sound speed in the compressed
layer. The derived quantities depend weakly on $\dot{\cal
N}_{\rm{\tiny{Lyc}}}$, somewhat on $n_{\tiny{{0}}}$, and strongly on
$a_{\rm{\tiny{S}}}$.

Whitworth et al.\ point out that the adopted value of
0.2~km~s$^{-1}$ for $a_{\rm{\tiny{S}}}$ is likely to represent a
lower limit of the sound speed in the layer, as both turbulence
generated by dynamical instabilities and extra heating from intense
sub-Lyman-continuum photons leaking from the \HII\ region tend to
increase this value. This point is very important because the mass
of the fragments depends strongly on this velocity:
$M_{\rm{fragment}} \propto a_{\rm{\tiny{S}}}^{40/11}$. Estimating an
accurate value of $a_{\rm{\tiny{S}}}$ in the hot PDR surrounding an
\HII\ region is an important issue, especially for a realistic
comparison with analytical models. In the following we will consider
values in the range 0.2--0.6~km~s$^{-1}$.

RCW~79 appears as an ``isolated'' \HII\ region, as observed in large
scale mid-IR images (MSX and Spitzer). The $^{13}$CO integrated
intensity map indicates that dense material, with densities of
$10^{3}$ cm$^{-3}$, is present in condensations at the periphery of
the ionized region, with weak emission outside this structure (see
Saito et al. \cite{sai01}). In the following we will consider
densities in the range 300--3000~cm$^{-3}$.

The rate of Lyman continuum photon emission by the exciting star of
RCW 79 is 2.9~$\times~10^{49}$~photons~s$^{-1}$ (Sect.~2). We adopt
this value in the following.

According to Whitworth et al., the time at which the fragmentation
starts is
$$t_\mathrm{frag} \approx
1.42\,\,a_{\rm{.2}}^{7/11}\,\,n_{3}^{-5/11}\,\mathrm{Myr}\, ,$$

the radius of the ionized region is then
$$
R_\mathrm{frag} \approx
6.4\,\,a_{\rm{.2}}^{4/11}\,\,n_{3}^{-6/11}\,\mathrm{pc}\, ,$$

and the mass of the fragments is
$$ M_\mathrm{frag} \approx 20.9 \,a_{.2}^{40/11}\,\,n_{3}^{-5/11}\, M_\odot\, ,$$
where $a_{\rm{.2}} = a_{\rm s}/0.2$ and $n_{3} = n_{\tiny 0}/1000$
are dimensionless variables.

The actual radius of the \HII\ region, $R\sim$~6.4~pc, allows us to
estimate its dynamical age, $t_\mathrm{dyn}$, which depends of the
density of the medium into which the region evolves. According to
Osterbrock~(\cite{ost89}), a star emitting 2.9~$\times~10^{49}$
ionizing photons per second forms a Str{\"o}mgren sphere of radius
$R_{0}\,\,(\mathrm{pc})~=~0.965\, \,n_{3}^{-2/3}$. This ionized
region expands; according to Dyson \& Williams~(\cite{dys97}), its
radius varies with time as
$$R = R_{0}(1 + 2.15\times 10^{-5}~t_\mathrm{dyn}/R_{0})^\frac{4}{7},$$
where $R$ and $R_{0}$ are in parsecs and $t_\mathrm{dyn}$ in years.

\begin{figure}[tb]
  \includegraphics[angle=0,width=85mm ]{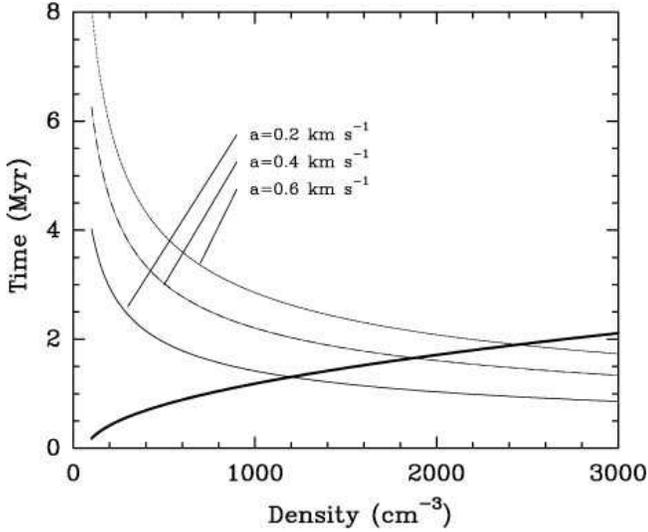}
  \caption{Variations of $t_\mathrm{dyn}$ (thick curve) and $t_\mathrm{frag}$ (thin curves) as a function of the density, $n_{\tiny{{0}}}$, for the observed radius of
6.4~pc and three values of $a_{\rm{\tiny{S}}}$ (0.2, 0.4, and
0.6~km~s$^{-1}$), respectively }
  \label{timedens}
\end{figure}

\begin{figure}[tb]
  \includegraphics[angle=0,width=85mm ]{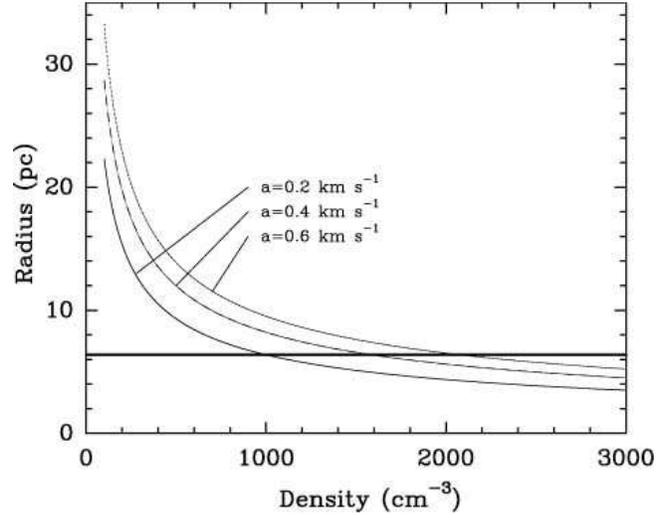}
  \caption{Radius of the layer at the time of fragmentation (thin curves), $r_\mathrm{frag}$, as a function
  of the density, $n_{\tiny{{0}}}$. The actual radius of RCW~79, 6.4~pc is shown (thick
  line)}
  \label{radiusdens}
\end{figure}

\begin{figure}[tb]
  \includegraphics[angle=0,width=85mm ]{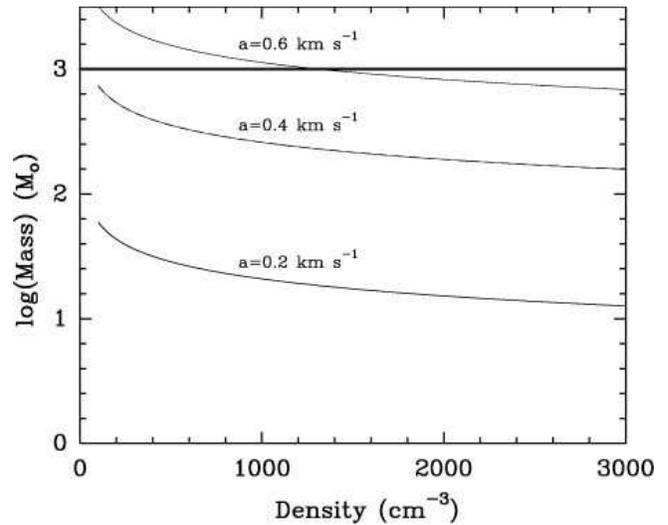}
  \caption{Mass of the fragments (thin curves) predicted by the model compared with the mass of the most massive observed
condensation, as a function of density, $n_{\tiny{{0}}}$. The mean
mass of the most massive fragment ($\simeq$1000\,$M_{\sun}$) is
shown (thick line) }
  \label{massdens}
\end{figure}

Fig.~\ref{timedens} shows how $t_\mathrm{dyn}$ and $t_\mathrm{frag}$
vary as a function of $n_{\tiny{{0}}}$, for three values of
$a_{\rm{\tiny{S}}}$ (0.2, 0.4, and 0.6~km~s$^{-1}$). Of course,
$t_\mathrm{dyn}$ must be larger than $t_\mathrm{frag}$, as we are
seeing molecular fragments already formed at the border of RCW~79.
This indicates that $n_{\tiny{{0}}} \geq 1200$~cm$^{-3}$ if
$a_{\rm{\tiny{S}}}$= 0.2~km~s$^{-1}$, or $n_{\tiny{{0}}} \geq
2400$~cm$^{-3}$ if $a_{\rm{\tiny{S}}}$=~0.6~km~s$^{-1}$.
In the same way Fig.~\ref{radiusdens} shows the radius of the layer
at the time of fragmentation, $r_\mathrm{frag}$, as a function of
$n_{\tiny{{0}}}$. This can be compared with the actual radius of the
\HII\ region, which must be larger than $r_\mathrm{frag}$ as
fragmentation has occured. This shows that $n_{\tiny{{0}}} \geq
950$~cm$^{-3}$ if $a_{\rm{\tiny{S}}}$= 0.2~km~s$^{-1}$, or
$n_{\tiny{{0}}} \geq 2000$~cm$^{-3}$ if $a_{\rm{\tiny{S}}}$=
0.6~km~s$^{-1}$.
Fig.~\ref{massdens} allows one to compare the mass of the fragments
estimated by the model and the mass of the most massive observed
condensation. {\it This comparison shows that massive fragments can
only form if} $a_{\rm{\tiny{S}}}$ {\it is high}.

Adopting a reasonable value of $n_{\tiny{{0}}} = 2000$~cm$^{-3}$,
we estimate a dynamical age of 1.7~Myr for RCW~79. According to the
model and assuming $a_{\rm{\tiny{S}}} = 0.4$~km~s$^{-1}$,
fragmentation occured some 10$^5$ years ago, after 1.6~Myr of
evolution; the radius of the \HII\ region was then 5.6~pc. This
constrains the age of the compact \HII\ region and its exciting
cluster. We adopt, for the radius of this region, 1.7~pc, which is
the radius of the 8\,$\mu$m PAHs emission ring surrounding it. A
dynamical age of 0.13~Myr is then estimated for the compact \HII\
region, assuming $n_{\tiny{{0}}}~=~3000$~cm$^{-3}$ for
$a_{\rm{\tiny{S}}} = 0.4$~km~s$^{-1}$. This result is
compatible with the general evolutionary scheme of RCW~79, in
particular with the high number of Class~I sources (of age of about
10$^5$ years) observed towards the most massive fragments. However
the model does not account for the large masses of fragments 2, 3
and 4 ($M_\mathrm{frag} \sim 200~M_{\sun}$ with the adopted figures;
see Fig.~\ref{massdens}). Note that adopting a higher value for the
sound speed (0.6~km~s$^{-1}$) and a higher density (2500~cm$^{-3}$)
lead to an unreasonably young dynamical age of 37000 years for the
compact \HII\ region.

The model of Whitworth et al. assumes expansion into an infinite and
homogeneous medium. This is probably unrealistic and may explain some of
the discrepancies between the predictions of the model and the
observations. A massive star forms in a dense core, but as the \HII\
region grows in size it probably expands into a medium of lower
density. Hence the necessity of models taking into account evolution
in a non-homogeneous medium, such as the models developed by Hosokawa
(private communication). Also, Whitworth et al. assume spherical
symmetry around the exciting star. Thus some fragments, and
subsequently some YSOs, should be observed in the direction of the
ionized gas. This is not the case: in RCW~79, as well as in Sh~104
(Deharveng et al.\ \cite{deh03}), the main fragments and the YSOs
seem to form in a preferential plane. Thus non spherical models are
also needed.

Other star-formation triggering mechanisms such as the
pressure-induced collapse of pre-existing clumps (Lefloch \&
Lazareff \cite{lef94}) and/or the growth of dynamical instabilities
in the collected layer (Garcia-Segura \& Franco \cite{gar96}) cannot
be ruled out in the case of RCW~79. In particular, the morphology of
the south-west ionization front, as traced by the GLIMPSE 8\,$\mu$m
filaments, is very chaotic, showing protruding structures (elephant
trunks or small-scale bright rims). The presence of such small-scale
structures is expected if the ionization front moves into an
inhomogeneous medium, or if small-scale clumps are formed on short
time scales by dynamical instabilities. For example, the formation
of object 6 near condensation 4 (Fig.~\ref{cond4smmf}), observed on
the top of  a bright rim, probably results from the pressure
induced-collapse of such a structure. However, the more massive
fragments observed along the annular collected layer, and especially
the most massive fragments (condensations 2 and 4), diametrically
opposite each other along this ring, point to a prevailing
large-scale, long time scale mechanism such as the collect and
collapse process.

\section{Conclusions}
%---------------------------------
We have presented a new 1.2-mm continuum map and near-IR images of
the Galactic \HII\ region RCW~79. The 1.2-mm map reveals the
presence of a layer of cold dust at the periphery of the ionized
region. This material has most probably been {\it collected}, during
the expansion of the \HII\ region, between the ionization and the
shock fronts. This layer is presently fragmented. Five large and
massive fragments are observed along the borders of RCW~79. The two
most massive fragments are diametrically opposite in this
layer.

The three most massive fragments are associated with young, massive
objects, as revealed by near- and mid-IR data. These objects are
mainly luminous Class~I sources. Some other are associated with
nebulosities with typical mid-IR colours of filaments in PDRs,
indicating that they are possibly early B stars surrounded by small
PDRs.

Signposts of recent star formation (maser emission associated with a
Class~I source) nearby a compact \HII\ region indicate that star
formation is still active there.

Kinematic information from H$\alpha$ observations reveals that
RCW~79 may experience a champagne phase acting as a destructor of
the surrounding annular structure. This fact complicates the
modeling of this region . However, it is probable that the
fragmentation of the annular structure occurred prior to the
champagne phase. The formation of massive fragments in the layer
favours the creation of lower density zones through which the
ionized gas can escape easily.

The analytical model of Whitworth et al.~\cite{whi94}, describing
the collect and collapse process, accounts for the global
properties of this region. RCW~79 is 1.7~Myr old; the collected
layer fragmented some 10$^5$~yrs ago; this is also the age of the
compact \HII\ region observed at the periphery of RCW~79, of its
exciting cluster and also probably of the numerous Class~I sources
observed towards the most massive fragments. The large masses of the
observed fragments indicate a large sound velocity (at least
0.4~km~s$^{-1}$) in the compressed layer; this is to be expected in
a hot PDR surrounding an \HII\ region. However the observations show
that the fragments and the massive YSOs are all found in a
preferential plane, which cannot be explained by Whitworth's model.
Non-spherical, non-homogeneous density models are needed.

Different processes of triggered star formation are probably
simultaneously at work in this region. Some YSOs are associated with
small-scale structures such as bright rims. Their formation has
probably been triggered by the pressure-induced collapse of a
pre-existing molecular clumps, or of clumps resulting from dynamical
instabilities in the collected layer. The large and massive
fragments observed at the periphery of RCW~79 most probably result
from the {\it gravitational collapse} of the layer of collected
material, according to the collect and collapse process.

All the new observational facts presented in this paper indicate
that we are dealing with the collect and collapse process as the
main triggering agent of {\it massive} star formation at the borders
of RCW~79. The presence of obscured zones towards the peaks of the
millimetre condensations indicate that precursor sites of
massive-star formation may still be present, representing ideal
sites to address the question of massive star formation.
%------------------------
\begin{acknowledgements}
%------------------------
We would like to thank R.~Cautain for his help in image processing,
and N.~Delarue and B. Boudey, students of the Universit\'e de
Provence, who worked on the presented data. R.~Zylka is thanked for
his help and advice at the beginning of the SIMBA data reduction. We
thank B. Lefloch for his collaboration in this long term project. We
would also like to thank Martin Cohen for providing the radio map of
the region. This research has made use of the Simbad astronomical
database operated at CDS, Strasbourg, France, and of the interactive
sky atlas Aladin (Bonnarel et al.~\cite{bon00}). This publication
uses data products from the Midcourse Space EXperiment, from the Two
Micron All Sky Survey and from the InfraRed Astronomical Satellite;
for these we have used the NASA/IPAC Infrared Science Archive, which
is operated by the Jet Propulsion Laboratory, California Institute
of Technology, under contract with the National Aeronautics and
Space Administration. We have also used the SuperCOSMOS survey. This
work is based in part on GLIMPSE data obtained with the Spitzer
Space Telescope, which is operated by the Jet Propulsion Laboratory,
California Institute of Technology under NASA contract 1407.

\end{acknowledgements}

%--------------------

%------------------------
{}
%______________________________________________________________________________
%______________________________________________________________________________

\end{document}